\documentclass[debug]{rmaa}
\usepackage{paralist}
\usepackage{psfrag,color}
\usepackage[latin1]{inputenc}

\def\msun{\mbox{M$_\odot$}}

\title{The delayed contribution of low and intermediate mass stars
  to chemical galactic enrichment: An analytical approach}

\author{
I. Franco\altaffilmark{1,2}
and L. Carigi\altaffilmark{1,3}
}

\altaffiltext{1}{Instituto de Astronom\'{\i}a, UNAM, M\'exico.}

\altaffiltext{2}{Max-Planck-Institut f\"ur Astronomie, Germany.}

\altaffiltext{3}{Centre for Astrophysics, UCLan, UK.}

\shortauthor{Franco \& Carigi}

\shorttitle{Delayed Contribution of LIMS}

\fulladdresses{
\item
Isabel Franco:
Max-Planck-Institut f\"ur Astronomie, Heidelberg, Germany
(franco@mpia-hd.mpg.de).
\item
Leticia Carigi: Instituto de Astronom\'{\i}a, Universidad Nacional Aut\'onoma de M\'exico,
Apdo. Postal 70-264, M\'exico 04510 D.F., M\'exico
(carigi@astroscu.unam.mx).
\item
Leticia Carigi:
Centre for Astrophysics, University of Central Lancashire,
Preston, Lancashire, PR1 2HE, United Kingdom
(lcarigi@uclan.ac.uk).
}

\listofauthors{I. Franco \& L. Carigi}
\indexauthor{Franco, I.}
\indexauthor{Carigi, L.}

\abstract{
We find a new analytical solution for the chemical evolution equations, 
taking into account the delayed contribution of all low and intermediate 
mass stars (LIMS) as one representative star that enriches the interstellar medium.  
This solution is built only for star formation rate proportional to the gas mass
in a closed box model. 
We obtain increasing C/O and N/O ratios with increasing O/H, 
behavior impossible to match with the Instantaneous Recycling Approximation (IRA).
Our results, obtained by two analytical equations, are very similar to those found by
numerical models that consider the lifetimes of each star.
This delayed model reproduces successfully the evolution of C/O$-$O/H and $Y-O$ relations
in the solar vicinity.
This analytical approximation is a useful tool to study the chemical
evolution of elements produced by LIMS when a galactic chemical evolutionary code
is not available.
}

\resumen{
Encontramos una nueva soluci\'on anal\'{\i}tica para las ecuaciones de 
evoluci\'on qu\'{\i}mica
tomando en cuenta la contribuci\'on retrasada de todas las estrellas de $m < 8$ \msun
(LIMS)
como una estrella representativa que enriquece al medio interestelar.
Esta soluci\'on es construida para tasa de formaci\'on estelar proporcional a la masa de gas
en un modelo de caja cerrada.
Obtenemos incrementos en C/O y N/O cuando O/H aumenta,
comportamiento imposible de igualar con IRA.
Nuestros resultados, obtenidos por dos ecuaciones anal\'{\i}ticas, son muy similares
a aquellos encontrados por modelos num\'ericos que consideran el tiempo de vida de cada estrella.
Este modelo retrasado reproduce 
la evoluci\'on de C/O-O/H y $Y-O$ en la vecindad solar. 
Esta aproximaci\'on anal\'{\i}tica es una herramienta util para estudiar
la evoluci\'on de elementos producidos por las LIMS
cuando no se dispone de un c\'odigo de evoluci\'on qu\'{\i}mica. 
}

\addkeyword{galaxies: abundances}
\addkeyword{galaxies: evolution}
\addkeyword{ISM: abundances}

\begin{document}

\maketitle

\section{Introduction}
\label{sec:intro}

Chemical evolution models are used to describe the temporal variation of the
  gas mass and the abundances of the different chemical elements
  that are present in the gas.
  Their importance relies on the fact that the chemical  history of
  the studied object (i.e. interstellar medium  in  galaxies and intergalactic
  medium) can be inferred.
  Moreover it is possible to get chemical information about stellar
  population properties and characteristics of the galaxies that we observe in the low 
  redshift Universe.

  The chemical evolution equations, shown by Tinsley (1974) and corrected later
by Maeder (1992), are relatively complex and can
  be solved through numerical models. There are some analytical approximations
  that have the advantage of
  predicting the general behavior of chemical elements in a quick and easy way
  but some precision may be lost.

One of the most well known analytical approximations is the Instantaneous
  Recycling Approximation (IRA) (Talbot \& Arnett 1971) where the star
  lifetimes are negligible compared with the age of the galaxies. This approximation
  has been widely used because it simplifies the solution to the chemical
  evolution equations, however, massive star (MS) lifetimes are on the order of $10^{6} - 
10^{7}$ years while the lifetimes for
  the low and intermediate mass stars (LIMS) are on the order of $10^{8} - 10^{10}$
  years comparable to the lifetime of a galaxy. Hence, IRA provides only a very rough 
approximation for elements produced by LIMS.

Serrano \& Peimbert (1983) proposed an analytic approximation related to the delays in chemical
enrichment in N/O-O/H relation assuming N and O yields increase with $Z$.
 They present closed and open models (with and without gas flows, respectively) that takes into
account the delay on nitrogen production due to LIMS concluding that nitrogen must be an element mainly secondary.

Later on, Pagel (1989) presented another approximation for  open models
 introducing an arbitrary time delay in order to study the chemical evolution
  of element produced by SNIa and LIMS, such as Fe and Ba (through s-process).
This time delay term makes the
  stars release the processed material to the interstellar medium (ISM) at a single time after the star formation.
After this single time delay, the contribution of all type of star is instantaneous,
like IRA.

  The objective of this work is to present an alternative analytical solution
  to the chemical evolution equations that considers the LIMS lifetimes as only 
  one group 
 with  delay times during the whole evolution. Those delay times are different
for each chemical element and are computed based on the characteristics of the
stellar yields and the stellar population.
 The delayed contribution to the
  chemical enrichment of elements produced by this type of star at different
  times should give results with a precision intermediate between the results
  obtained by using IRA and the ones obtained by numerical codes. As an
  application, this work aims to reproduce the C/O \textit{vs} O/H and $Y(O)$ histories
  indicated by the stars and HII regions at the solar vicinity.

This approximation is a simple tool for theoretical and observational astronomers
who need to include chemical aspects in their computations or to
interpret observational data when they do not have access to
a numerical code of chemical evolution of galaxies.
Our equation for the mass abundance of element $i$, $X_i(t)$, 
would replace the popular equation of $Z(t)$
obtained assuming IRA in the closed box regimen, erroneously applied
for elements produced by LIMS (e.g. He, C, N).

  \section{The delayed contribution model}
\label{sec:models}

  This approximation finds an analytical solution to the chemical 
evolution
  equations.
  In the context of a closed box model, we take into account the stellar 
lifetimes ($\tau$) of the LIMS as a group where each of its members die at the same time.  
  In this approximation, after a star forming burst, all massive stars (MS) with initial mass $m$ higher 
than $8 M_{\odot}$ enrich the interstellar 
medium (ISM) instantaneously while every LIMS with $m < 8 M_{\odot}$ 
enriches the ISM after one single time $\tau_L$, for the total mass ejected.
   
  We divide $R$, the material returned by the stars to the ISM, in two terms: $R_{L}$ and $R_{M}$, the returned mass by 
the LIMS and the MS, respectively:

$$R_{L}=\int_{M_{1}}^{8M_{\odot}}{(m-m_{R})\phi(m)dm}$$
  
$$R_{M}=\int_{8M_{\odot}}^{Msup}{(m-m_{R})\phi(m)dm}$$ 

\noindent  
 where $m_{R}$ is the remnant mass, $M_{1}$ and $M_{sup}$ are the lowest and highest 
star that contributes to the ISM,and $\phi$ is the initial mass function.

Taking the galactic chemical evolution equations for the gas mass in a closed box model

$$\frac{dM_{gas}}{dt}=-\Psi(t)+E(t)$$

  \noindent
  where $\Psi(t)$ is the star formation rate and $E(t)$
  is the gas rate that is returned to the ISM from stars that die.
  In our approximation, we have divided $E(t)$
  in two parts, the rate where all MS return material to the ISM,
  regarding that, $t - \tau(m) \approx t $:

$$E_{M}(t)=\int_{8 M_{\odot}}^{Msup}{(m-m_{R})\Psi(t-\tau(m))\phi(m) dm}=R_{M}\Psi(t)$$ 

\noindent
and the rate where the LIMS return material to the ISM, $t - \tau(m) \approx t - 
  \tau_L$,so that:

$$E_{L}(t)=\int_{M_{1}}^{8M_{\odot}}{(m-m_{R})\Psi(t-\tau(m)) \phi(m)
  dm}=R_{L} \Psi(t-\tau_L)$$

  We have considered that the star formation rate is proportional to the 
  gas mass with a constant efficiency $\nu$, $\Psi(t)=\nu M_{gas}(t)$, therefore,

$$\frac{dM_{gas}}{dt}=-\nu (1-R_{M}) M_{gas}(t) + \nu R_{L}M_{gas}(t-\tau_L)$$

  The first term of this equation represents the contribution of the MS 
  while the second one represents the delayed contribution of the LIMS.

  In order to get the abundance evolution equation, we 
  divide the stellar population yields $P_{i}$ in two parts as we did 
  with $R$, $P_{i}=P_{L_i}+P_{M_i}$, with $P_{L_i}$ and $P_{M_i}$ defined by:  

$$P_{L_{i}} =\int_{M_{1}}^{8M_{\odot}}{mp_{i}(m)\phi(m)dm}$$

$$P_{M_{i}} =\int_{8M_{\odot}}^{Msup}{mp_{i}(m)\phi(m)dm}$$

\noindent
where $p_{i}(m)$ represents the stellar yield of element $i$, that is the fraction of
  initial stellar mass processed and ejected by a star of initial mass $m$.

  Taking the galactic chemical evolution equation for the mass of gas in 
  element $i$, $ F_{i}(t)=X_{i}(t) M_{gas}(t)$, in a closed box model,
 
  $$\frac{dF_{i}}{dt}=-X_{i}(t)\Psi(t)+E_{i}(t)$$

  Dividing the $E_{i}$ in two parts as we did before with $E(t)$ we get,
  
$$E_{{M}_i}(t)=\int_{8M_{\odot}}^{Msup}{[(m-m_{R})X_{i}(t-\tau(m))+mp_{i}(m)]
  \Psi(t-\tau(m))\phi(m)dm}$$
  
$$E_{{M}_{i}}(t)=[R_{M}X_{i}(t)+P_{{M}_{i}}] \Psi(t)$$

$$E_{{M}_{i}}(t)=\nu R_{M}F_{i}(t)+\nu P_{{M}{i}}M_{gas}(t)$$
 
  And for the LIMS:

$$E_{L_i}(t)=\int_{M_1}^{8M_\odot}{[(m-m_R)X_i(t-\tau(m))+mp_i(m)]\Psi(t-\tau(m))\phi(m)dm}$$

$$E_{{L}_{i}}(t)=[R_{L}X_{i}(t-\tau_{i})+P_{{L}_{i}}]\Psi(t-\tau_{i})$$
  
$$E_{{L}_{i}}(t)=\nu R_{L}F_{i}(t-\tau_{i})+\nu P_{{L}_{i}}M_{gas}(t-\tau_{i})$$

  We finally obtain:
   \begin{eqnarray}
   \frac{dF_{i}}{dt}=-\nu (1-R_{M})F_{i}(t)+\nu P_{M_{i}}M_{gas}(t)+ \nu R_{L}F_{i}(t-\tau_{i}) 
   +\nu P_{L_{i}}M_{gas}(t-\tau_{i}) \nonumber 
   \end{eqnarray}

  For each element $i$, the group of LIMS enrich the ISM at a time
  $\tau_{i}$ after it was formed; $\tau_{i}$ has to be chosen as the representative ejecting 
  time of the population.
  Equivalently, a representative mass is also required: this is defined as the mass up to which 
  the accumulated chemical yield of stars of lower mass is equal to the half of the entire 
  yield of the LIMS population. As the stellar yields are different for each chemical element and 
  strongly depend on the metallicity, we calculate the value $M_{repr}$ for each chemical element 
  and for each metallicity:
  
 $$\int_{M_{1}}^{M_{repr}}{mp_{i}\phi(m)dm}=\frac{1}{2}\int_{M_{1}}^{8M_{\odot}}{mp_{i}\phi(m)dm}$$
  
  \section{Solution to the Chemical Evolution Equations}
\label{sec:solutions}

  We solve the differential equations for $M_{gas}$ and $F_i$
  assuming for the moment that the metallicity of the stars that enrich the ISM is constant in the evolution, 
  therefore the stellar properties, $\tau$, $R_{M}$, $P_{M}$, $R_{L}$,$P_{L}$ are constant.

  \subsection{Gas Mass  Evolution Equation}

  The equation to be solved is:
  \begin{equation}
  \frac{dM_{gas}}{dt}=-\nu (1-R_{M}) M_{gas}(t) + \nu R_{L}M_{gas}(t-\tau_L)
  \end{equation}

  For $t<\tau_L$, the equation can be solved by IRA, which gives:

  $$M_{gas}(t)=M_{gas}(0) e^{-\xi t} \qquad 0<t<\tau_L,$$
  
\noindent
  where $\xi= \nu (1-R_{M})$.

  Having solved for the first interval, $0<t<\tau_L$, we now substitute this solution in (1) 
  when $\tau_L < t < 2\tau_L$ obtaining as solution:

$$M_{gas}(t) = M_{gas}(0)[e^{-\xi t}+e^{-\xi (t-\tau_L)}(t-\tau_L)(\nu R_{L})]$$
  
  Solving for the other intervals:

  \begin{equation}\label{ru4} 
  M_{gas}(t)= M_{gas}(0) e^{-\xi t} \sum_{k=0}^{n-1}{\frac{[e^{\xi \tau_L} (t-k\tau_L)(\nu R_{L})]^k}{k!}} 
  \end{equation}

\noindent
  for $(n-1)\tau_L< t< n\tau_{L}$.

  Taking into account the delayed contribution due to the LIMS, the gas mass equation is
  the product of the term that comes from the contribution of the MS (IRA) times by a summation
  that represents the contribution of LIMS that were born at $t-k\tau_L$ and enriched the ISM at the time $t$.

  \subsection{Gas Mass in Element $i$}

  The equation to solve is:

  \begin{eqnarray}
  \frac{dF_{i}}{dt}=-\nu (1-R_{M})F_{i}(t)+\nu P_{M_{i}}M_{gas}(t)+ \nu 
   R_LF_{i}(t-\tau_{i})+ 
  \nu P_{L_{i}}M_{gas}(t-\tau_{i})
  \end{eqnarray}
 
  Following a similar procedure to that of previous section:

  $$F_{i}(t)=M_{gas}(0)e^{-\xi t} [X_i(0)+\nu P_{M_{i}}t] \qquad 0<t<\tau $$

  Solving for the second interval:

  \begin{eqnarray} 
  F_{i}(t)=M_{gas}(0)[e^{-\xi t} [X_i(0)+\nu P_{M_{i}}t]+e^{-\xi (t-\tau_i)} 
  (t-\tau_{i})
  (\nu R_{L})[X_i(0)+\nu P_{M_{i}}(t-\tau_{i})+P_{L_{i}}/ R_{L}]] \nonumber 
  \end{eqnarray}
 
\noindent
  for $\tau_{i}<t<2\tau_{i}$.

Solving for the other intervals:

\begin{equation}
F_i(t)=M_{gas}(0)e^{-\xi t} \sum_{k=0}^{n-1}
{\frac{[e^{\xi \tau_i}(t-k\tau_i)(\nu R_L)]^k}{k!}} \times
{([X_{i}(0)+\nu P_{M_{i}}(t-k\tau_{i})] + kP_{L_{i}}/R_L)}
\end{equation}

\noindent
for $ (n-1)\tau_i < t < n \tau_i$.

  \subsection{Chemical Abundances by Mass $X_{i}(t)$}

  The evolution equation for $X_{i}(t)$ is obtained from the ratio
  of the mass in the form of element $i$ and the gas mass:

$$X_{i}(t)= \frac{F_{i}(t)}{M_{gas}(t)}$$

For the total gas mass  we assume that all the LIMS population ejects the material in 
  form of element i at the same time as the material with all the 
  elements, i.e., $\tau_{i}=\tau_L$.
Then we finally get $X_{i}(t)$
  dividing eq. (4) by eq. (2) where the abundance ratios are 
  independent of the initial value of gas mass, $M_{gas}(0)$.
Therefore

\begin{equation}
X_{i}(t)= \frac{\sum_{k=0}^{n-1} G(k,t) Q(k,t)}{\sum_{k=0}^{n-1} G(k,t)}
\end{equation}

\noindent
where
$ G(k,t)=\frac{[(t-k\tau_i)(e^{\xi \tau_i} \nu R_{L})]^k}{k!} $
and
$ Q(k,t)=\nu P_{M_i}(t-k\tau_{i}) + X_i(0)+kP_{L_i}/R_L$,
valid for $t - k\tau_i > 0 $.

  To solve the approximation of  $\tau_{i}=\tau_L$ in $M_{gas}(t)$, we
  take the gas mass as the average of the obtained masses of each chemical
  element:

$$  M_{gas}(t)=\frac{\sum_i{M_{gas}(t,\tau_i)}}{5} $$  

\noindent
  as the gas mass is independent of the element $i$, and 5 elements
are considered here (He, C, N, O, and $Z$).

  We have found that the values of $M_{gas}(t)$ and each $M_{gas}(t,\tau_{i})$ are  
  quite similar in values except for high $\nu$ and long $\tau_i$.
  In such cases, $X_{i}(t)$ show an artificial secondary raise (see Section 4 for details).

  \subsection{Stellar Properties}

  We have chosen an age, an initial mass function, and a yield set in order to study the applications and
  limits of our approximation.We consider 13 Gyr as the age of the models, the time elapsed since the beginning 
  of the formation.
  The Initial Mass Function adopted is the one proposed by Kroupa, Tout \& Gilmore (1993)
  in the mass interval given by 0.1 $< m/M_{\odot} <$ 80, hence, $M_{sup}=80 M_{\odot}$ and $M_{1}=1 M_{\odot}$,the canonical lowest mass star that enriches the ISM in the IRA approximation. 

  After trying several stellar yields published in the literature, we selected those that  provide us with the most complete information in terms of mass, metallicities and physical parameters such as stellar rotation, mass loss due to stellar winds and elements produced during the HBB stage.
  Once they were selected, we calculated the yields of the population based on its initial metallicity. 

  For LIMS we use the stellar yields and remnants provided by van den Hoek \& Groenewegen (1997)
  that take in account the mass loss due to the stellar winds. These yields are characterized 
  by the  parameter $\eta$ that represents the mass loss, we chose a range for $\eta$ for 
$Z_{pop}=0.001$, 0.004 and 0.02.
  In order to find the delay times of the LIMS group, we have adopted the stellar lifetime by Schaller (1992).
  We have assumed that no LIMS formed with $Z_{pop} \sim 0$.

  For MS we use the stellar yields and remnants provided by Hirschi (2007) 
  for $Z_{pop}=10^{-8}\sim 0$, Meynet \& Maeder (2002) for  $Z_{pop}=10^{-5}$ and 0.004,
  and Maeder (1992) for $Z_{pop}=0.02$.
  We interpolate linearly  by mass the stellar yields and remnant
  mass of LIMS and MS, we also extrapolate linearly in mass in order to reach $m_{up} = 80 M_{\odot}$. 

  We do not interpolate the stellar properties by metallicity and we combine MS and LIMS properties according to 
  the initial stellar metallicity.
For LIMS of $Z_{pop}=10^{-5}$ we assumed the $Z_{pop} = 0.001$ yields and remnants.

  Below we show the Tables 1 and 2 with physical properties derived here for our approximation, 
  $R_M$, $R_L$, $P_{M_{i}}$, $P_{L_{i}}$, and $\tau_i$.
  As you can notice, all values of $\tau_{i}$ are $<1$ Gyr, with the exception of $\tau_O$ 
  for $Z_{pop}=0.02$, which is 5.89 Gyr. 
  This difference is important because it causes an artificial behaviour in oxygen (only for 
  the case with very high $\nu$) and therefore in all abundance ratios related to oxygen.

 \begin{table}[h]\centering
\setlength{\tabnotewidth}{0.5\columnwidth}
  \tablecols{5}
\setlength{\tabcolsep}{0.5\tabcolsep}
\caption{Returned masses, stellar population yields, representative masses (in $M_{\odot}$)
and delay times (in Gyr)
for low and intermediate mass stars.}
\label{lims}
\begin{tabular}{lcccc} \toprule
  $Z_{pop}$  &  $\sim$ 0   & 0.001 & 0.004 & 0.02\\ 
\midrule
$R_L$                 & --- &   0.217&  0.225&  0.238\\ 
\hline
$P_{He}\times10^{-2}$ & --- &   1.261&  1.035&  0.801\\ 
$M_{repr}^{He}$       & --- &   2.270&  2.414&  2.646\\ 
$\tau_{He}$           & --- &   0.719&  0.635&  0.623\\ 
\hline
$P_{C} \times10^{-3}$ & --- &   2.680&  1.717&  0.585\\
$M_{repr}^C$          & --- &   2.022&  2.223&  2.761\\ 
$\tau_{C}$            & --- &   0.968&  0.788&  0.554\\ 
\hline
$P_{N} \times10^{-4}$ & --- &  11.846&  9.300&  8.162\\ 
$M_{repr}^N$          & --- &   4.769&  5.094&  4.769\\ 
$\tau_{N}$            & --- &   0.110&  0.099&  0.121\\ 
\hline
$P_{O}\times10^{-4}$  & --- &   2.937&  1.571&  0.790\\ 
$M_{repr}^O$          & --- &   2.298&  2.362&  1.185\\ 
$\tau_O$              & --- &   0.697&  0.672&  5.890\\ 
\hline
$P_{Z}\times10^{-3}$  & --- &   4.206&  2.832&  1.492\\ 
$M_{repr}^Z $         & --- &   2.546&  2.779&  3.126\\ 
$\tau_Z$              & --- &   0.532&  0.432&  0.390\\ 
\bottomrule
\end{tabular}
\end{table}

\begin{table}[h]\centering 
\setlength{\tabnotewidth}{0.5\columnwidth}
  \tablecols{5}
\setlength{\tabcolsep}{0.5\tabcolsep}
\caption{Returned masses and stellar population yields for massive stars.}
\label{ms}
\begin{tabular}{lcccc} \toprule
$Z_{pop}$ & $\sim$ 0  & $1.0\times10^{-5}$ & 0.004 & Z=0.02\\ 
\midrule
$R_M \times 10^{-2}$     & 7.345& 7.339& 7.370& 7.438\\ 
$P_{He}\times 10^{-2}$   & 1.208& 1.212& 1.092& 0.929\\ 
$P_{C}\times 10^{-3}$    & 2.694& 0.854& 0.820& 2.802\\ 
$P_{N}\times 10^{-4}$    & 2.363& 0.429& 1.333& 4.689\\ 
$P_{O}\times 10^{-3}$    & 6.261& 6.104& 7.285& 2.953\\ 
$P_{Z}\times 10^{-2}$    & 0.988& 1.012& 1.154& 0.893\\ 
\bottomrule
\end{tabular}
\end{table}

  \section{Results}

  We study the evolution of mass abundances H, He, C, N, O and $Z$, using the delayed contribution model
  taking into account primordial abundances, $X(0)= 0.75$, $Y(0)= 0.25$  for all metallicities,
  and we follow the evolution from $t=0$ until $t=13.0$ Gyr with a time step of $\Delta t=0.01$ Gyr.
  In this section we do not show the evolution for $Z_{pop} \sim 0$, since we assume that all Pop 
  III stars are MS so that, C/O(t) and N/O(t) are constant according to IRA.

  Next we discuss the result obtained when using our model for different values of
  the gas consumptions $\mu=M_{gas}/M_{tot}$ as well as for different metallicities.

  \subsection{Evolution of C/O \textit{vs} O/H}

  In Fig.1, we show the evolution of C/O vs O/H for three different metallicities and three $\mu$ values.
  In each model the value of $\mu$ is reached at $t_g = 13.0$ Gyr.
The horizontal lines represent 
  the results assuming IRA, the lower lines when only MS are considered and the higher lines 
  when both MS and LIMS are considered.
Assuming IRA, the abundance ratios by number can be written as,
C/O(t)$=\frac{P_C/12.0}{P_O/16.0}$
  and O/H(t)$=\frac{-P_O ln\mu/16.0}{0.75 +(P_{He}+P_Z)ln\mu}$.

  Initially, O/H depends on the star formation rate (SFR), the smaller the gas consumption $\mu$ 
  the higher the O/H abundance value.

  Massive stars are the first that die and so the C/O values obtained with the delayed
  model are identical to the ones obtained using IRA considering only  massive stars.
  The effect of  delay can be seen when the curve begins to increase its slope making
  the abundance ratio C/O grow. Afterwards, the curve flattens coinciding with the IRA
  case considering MS and LIMS together.

  \subsubsection{Case $\mu = 0.1$}

  In this case 90 \% of the initial gas mass has become stars at the end
  of the evolution and only 10 \% remains as ISM (See Fig. 1, Panel a, b, c).

  When the metallicity has values of $Z_{pop} = 10^{-5}$ and 0.004,
  C/O increases for the first time when 12 + log(O/H) $\sim 8.0$ dex
  respectively (Panels a and b) that means in the times equal to $0.97$ Gyr and $0.79$
  Gyr that are the delay times for carbon ejection from LIMS for this metallicities (See
  Table 1).

  When $Z_{pop} = 0.02$, C/O is higher than the lower $Z_{pop}$ cases because $P_C/P_O$ are higher for MS
  at solar metallicity.
Moreover when $Z_{pop} = 0.02$ the values for O/H are smaller than
  those for other metallicities at the same times.
Both effects are due to MS of high $Z$,
  the metal rich massive star ejects more carbon in the stellar wind stage,
  leaving less carbon behind to be processed in order to become oxygen.

In the other hand, C yields of LIMS are lower
  at high $Z$, again for the mass loss rate; stellar winds are intense in LIMS and stars have
  less mass to produce heavy elements.
This makes less difference between the C/O ratios with IRA for MS and MS + LIMS.

  Moreover, for $Z_{pop}=0.02$ the first increment of C/O comes when 12+log(O/H) $\sim 7.5$ dex (Panel c)
  corresponding to a time delay for carbon of 0.55 Gyr.
  The curve keeps growing due to the increase of carbon relative to oxygen.
  Afterwards, the second increase is artificial (at 12+log(O/H)$\sim 8.6$ dex) and it is produced by 
  the oxygen dilution due to LIMS.
In the $\mu = 0.1$ case  the SFR is a quite  decreasing exponential function,
therefore for $t > \tau_O$ the number of LIMS that are diluting is higher than the number
 of MS that
  are enriching of O the ISM .

  \subsubsection{Case $\mu = 0.5$}

  In this case  50 \% of the initial gas mass has become stars at the end
  of the evolution and  50 \% remains as ISM, therefore the SFR is lower and flatter than in the 
  case $\mu = 0.1$.
  As a consequence the values for O/H are smaller in this case than what was previously found at 
  the same times (see Fig 1, Panels d,e,f).
  In the other hand, the C/O ratios obtained in this case are almost the same as for $\mu = 
  0.1$ because $C/O \propto P_C/P_O$, therefore C/O depends on the stellar yields not on $\mu$.

  The oxygen dilution effect is present again for $Z_{pop} = 0.02$ but to a smaller degree than for $\mu = 
  0.1$ due to the SFR behaviour.
  In the $\mu = 0.5$ fort $t > \tau_O$ the SFR is high so there are  many  MS enriching the gas making 
  difficult the oxygen dilution by LIMS.

  \subsubsection{Case $\mu = 0.9$}

  In this case only 10 \% of the initial gas mass has become stars at the end of the evolution and  90\% 
  remains as ISM, therefore the SFR is even lower and practically flat.

  In Panels (g,h,i) the results for this case are shown. This is the lowest SFR that we have studied, 
  which implies that O/H ratio is very low too. 
  In this case there is no second significant increase in C/O vs O/H,
  because the SFH is the flattest one and therefore the amounts of massive stars and LIMS are
 almost constant during the whole evolution.

\begin{figure}[!t]
\includegraphics[width=\columnwidth]{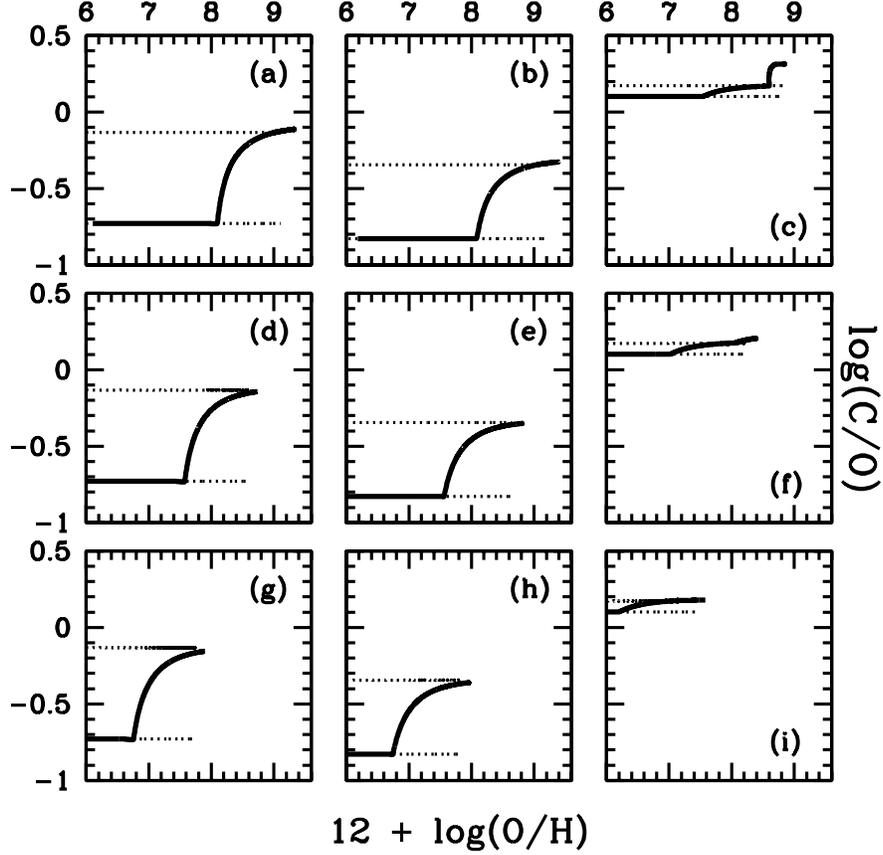}
\caption{Evolution of C/O abundance ratio versus O/H abundance ratio for metallicities
of the stellar population
  $Z_{pop}=10^{-5}$ (a, d, g), $Z_{pop}=0.004$ (b, e, h), $Z_{pop}=0.02$ (c, f, i)
  and three different values of gas consumption $\mu=0.1$ (a, b, c), $\mu=0.5$ (d, e, f) ,
  $\mu=0.9$ (g, h, i).
The dotted lines represent IRA results
when only MS (lower lines) or both MS and LIMS (upper lines) are considered.
}
  \label{co} 
  \end{figure}

  \subsection{Evolution of N/O \textit{vs} O/H}

 In Fig. 2 we show the N/O-O/H evolution for the same $\mu$ and $Z_{pop}$ values as in the previous 
 section.

 The general behavior of N/O ratio is similar to the C/O ratio because N and C are produced by MS and 
 LIMS and O is produced by MS mainly. Therefore when O/H ratio increases N/O ratio tends to increase 
 when the LIMS eject  material at the delay time.
 Since MS produce much less N than C, N/O is lower than C/O for early times.
 Also N/O evolution presents an artificial raise due to O dilution, as in the C/O history.

  \begin{figure}[!t]
  \includegraphics[width=\columnwidth]{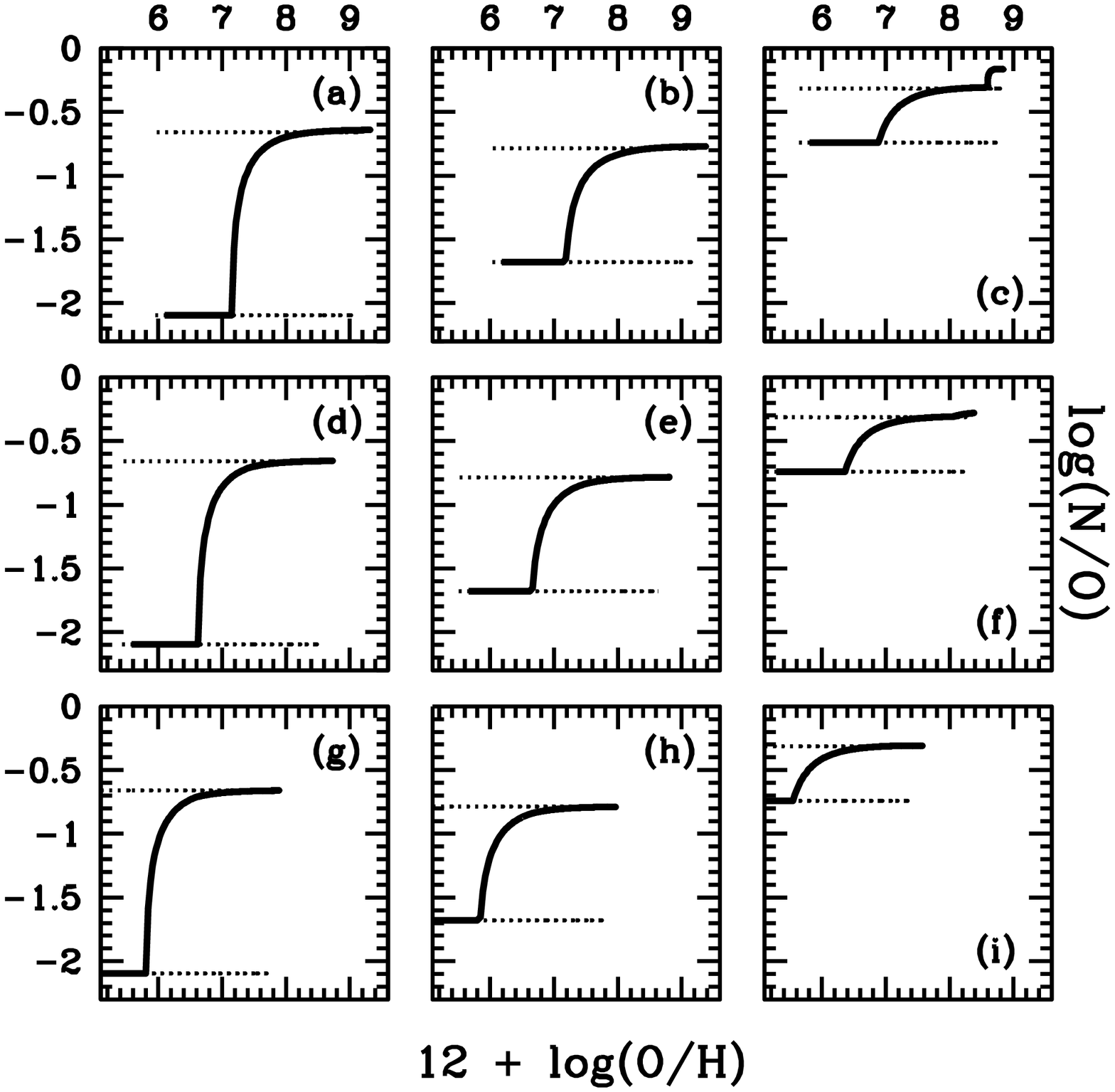}
  \caption{Evolution of N/O abundance ratio versus O/H abundance ratio for metallicities
of the stellar population
  $Z_{pop}=10^{-5}$, 0.004, and 0.02; and three different values of gas consumption
  $\mu = 0.1$, 0.5, and 0.9, as in Fig 1.}
  \label{no} 
  \end{figure}

%------------------------
  \subsection{Evolution of Helium with Oxygen}

  \begin{figure}[!t]
  \includegraphics[width=\columnwidth]{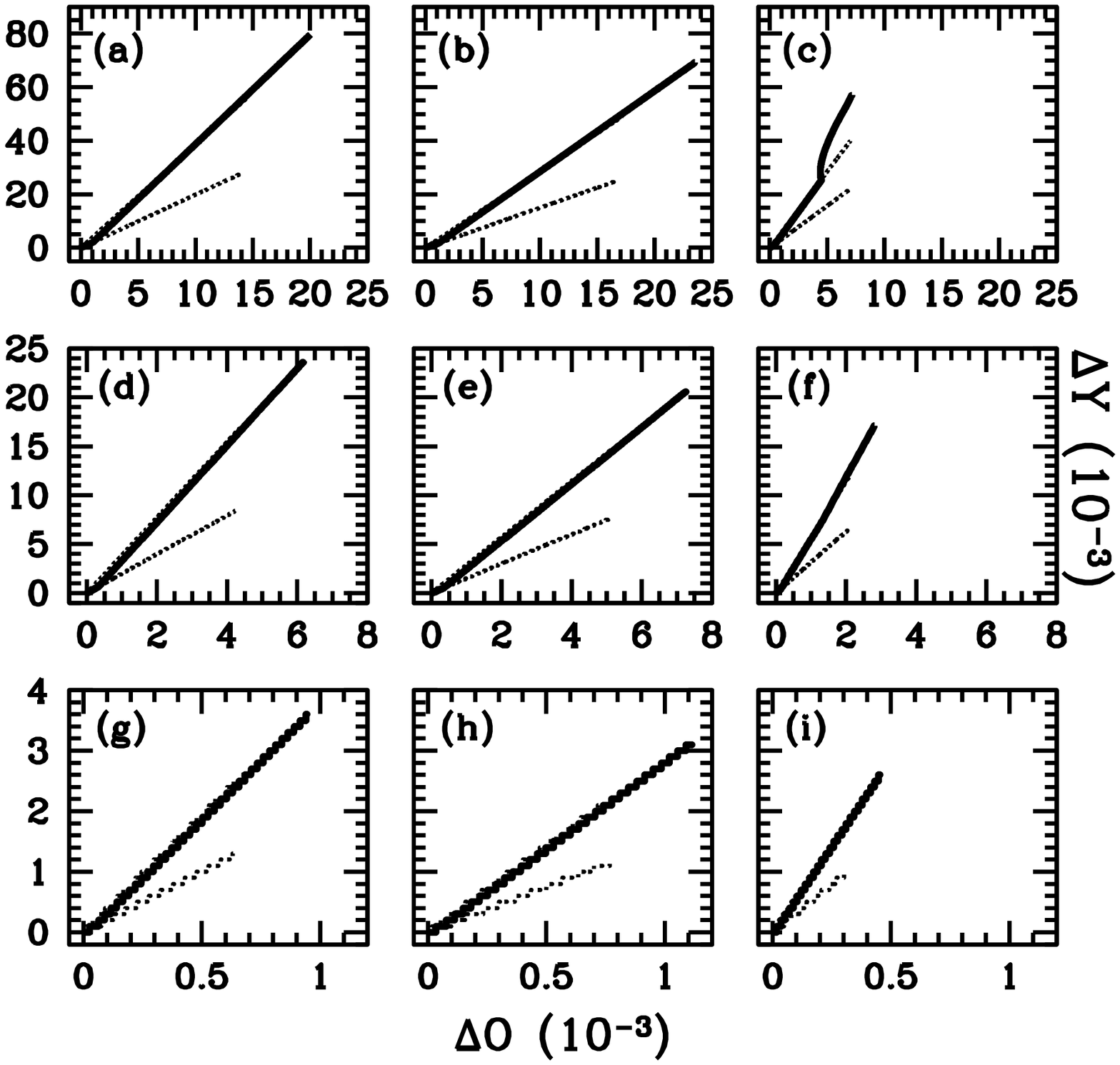}
  \caption{Evolution of Helium versus Oxygen, both by mass, for the metallicities 
of stellar population  $Z_{pop} = 10^{-5}$, 0.004 and 0.02 and three different values of gas consumption $\mu = 0.1$, 0.5 and 0.9,  as in Fig 1.
The dotted lines represent IRA results,
when only MS (flatter lines) or both MS and LIMS (steeper lines) are considered.
}
  \label{heo}
  \end{figure}

  In Fig. 3 we show the helium by mass ($Y$) vs oxygen also by mass ($O$)
 for the same $\mu$ and $Z_{pop}$ values as in sections 4.1 and 4.2.
  Here we have assumed for all $Z_{pop}$ that the initial $Y$ and $O$ values are 0.25 and 0.00, respectively.

  Since the delay time for LIMS is lower than 0.7 Gyr,
$\Delta Y$ values are very similar to those obtained assuming 
  IRA for MS and LIMS.
  The slopes of $Y(O)$ changes with $Z_{pop}$ due mainly to the $Z$ dependence of oxygen yields through the stellar winds.
  Also $Y(O)$ presents the second significant increase caused by the artificial oxygen dilution due the LIMS.

  \subsection{Limitation of the Approximation}

  Based on C/O-O/H and N/O-O/H and $Y-O$ evolutions the artificial second raise is present only 
  for $Z_{pop}=0.02$ and $\mu=0.1$.
  This second raise is due to  a huge delay time and a quite decreasing SFR.
  According to our approximation when we study O(t), 
LIMS eject oxygen and the rest of the elements at the same
  time, $\tau_{L}=\tau_{O}$.
  Since $P_{L_{O}}$ is very low, the rest of elements produced by LIMS dilute oxygen and
  there are not enough MS to counteract that dilution.

  This behavior indicates that $\tau_{L}=\tau_{i}$ approximation is not valid for an element $i$
mainly produce by MS, when the LIMS delayed contribution 
of the rest of elements is huge
compared to the MS contribution of the element $i$ at a fixed time.

%-----------------------------------------------------------------------------
  \section{The Solar Vicinity}

The solar vicinity  is the place in the Universe with the largest number of observations,
therefore we use it to test our approximation. However, this approximation was developed for a 
constant metallicity of the stars, therefore $P_{i}$, $\tau_{i}$ and $R_{i}$ are constant
during the gas evolution of a galaxy.
This assumption is not valid for a real galaxy, because the stars in a galaxy form with 
gas metallicity that changes in time. 
Therefore, the Figs. 1, 2, and 3 are good illustrative examples, but not real ones.

We compute a model where the stars are formed with metallicities similar to the gas metallicity when 
the SFR is ongoing.
We have made $M_{gas}$ continuous and $F_{i}$ when the gas reach the metallicities ($Z_{gas}$) of 
the stellar population ($Z_{pop}$) assumed in this work.

Formally, when  $Z_{pop}^j < Z_{gas} < Z_{pop}^{j+1}$ we assume that $P_i=\frac{P_i^j+P_i^{j+1}}{2}$
and $\tau_i=\frac{\tau_i^j+\tau_i^{j+1}}{2}$.
Specifically for $0 \leq Z_{gas} < 10^{-8}$ and $Z_{gas}>0.02$ we have assumed the stellar properties 
for $Z_{pop}=10^{-8}$ and $Z_{pop}=0.02$, respectively.
However, for $10^{-5} \leq Z_{gas} < 0.02$ we have considered  average stellar properties between two 
consecutive $Z_{pop}$'s.

With this combination we obtain more realistic results in the sense that we are taking into account 
the evolution of the stellar populations depending on its initial metallicity.
Another advantage of this combination is that different star formation rate efficiencies can be used
as a function of time representing in a more realistic way the formation of different components of a 
galaxy.

%-----------------------------------------------------------------------------
  \subsection{Observational Restrictions}

Now, we study the applicability and limitations of the delayed contribution model
with different $Z_{pop}$'s comparing our theoretical results with existing observations
of the solar vicinity.
We define the solar vicinity as the volume contained in a cylinder centered on the Sun with a radius of 
$\sim$ 1 kpc and height enough to reach objects located in the Galactic halo.

Since C is better known than N, we will test our delayed approximation with the C/O-O/H relation in 
the solar vicinity and our observational constraints for C and O are:

  \begin{itemize}
  \item{H II regions to test the model results at the present time
   (Esteban et al. 2005, Garc\'{\i}a-Rojas et al. 2004).}
  \item{Halo and disk main sequence stars at different times
   as past restrictions (Akerman et al. 2004).}
  \item{The Sun as a restriction at $4.5$ Gyrs ago (Asplund, Grevesse \& Sauval 2004).}
  \end{itemize}

  Since the Sun is located at $8$ kpc from the Galactic center,
  we used the two HII regions studied by Esteban et al. (2005) and Garc\'{\i}a-Rojas et al.
  (2004) nearest the Sun:
  Orion Nebula and NGC $3576$ at $r = 7.46$ kpc and $r = 8.40$ kpc, respectively.
The C/H and O/H values have been increased by 0.10 and
0.08 dex, respectively, owing to the fraction of C and O embedded in dust grains
 (Esteban et al. 1998).
  Akerman et al. (2004) show values for 34 F and G dwarf stars from the Galaxy halo
  combining them with similar data of 19 stars of the disk.

%-----------------------------------------------------------------------------
\subsection{ Chemical Evolution Models}

We have built four chemical evolution models to reproduce the O/H value in NGC3576.
Since our approximation was obtained for a closed box model with a SFR proportional to $M_{gas}$, 
all models presented in this section follow those assumptions and moreover consider the same IMF, 
mass range and stellar yields specified in section 3.4.

Each model is characterized by the approximation used in the lifetime.
Model 1 is our delay approximation, which assumes that the whole group of LIMS
is represented by a specific star and its lifetime is considered.
Models 2 and 3 are models in IRA, for MS only and MS and LIMS, respectively.
In those models no lifetime is considered.
Model 4 is obtained using CHEVO code (Carigi 1994) that considers the lifetime of each star until 
leaves the main sequence.

A closed box model with a $SFR = \nu M_{gas}$ that reproduces a final O value has only
one free parameter that is the efficiency $\nu$. Therefore,
the models 1, 2, 3 and 4 need   
 $\nu=$0.13, 0.13, 0.17, and 0.23 in order to get 12 + log(O/H) $\sim 8.82$, 
resulting in values of $\mu$ 0.29, 0.21, 0.21, and 0.51, respectively.

Since O is produced by MS mainly and Models 1 and 2 assume IRA for MS only
the SFR is identical for both models, but the value of $M_{gas}$ obtained by our delayed approximation
is higher than models with IRA due to the contribution of the representative mass of the LIMS as a 
group.
The SFR for Model 3 is slightly higher due to the dilution of O caused by LIMS at each time
and $M_{gas}$ are nearby identical because the high SFR in model 3 counteracts the material
returned to the ISM by LIMS.
In model 4, $\nu$ and $\mu$ values are higher than those obtained by the rest of the models because
CHEVO code needs a delay for each star that forms until its death enriching the ISM, therefore
the SFR have to be higher than the rest of models to reach the same O value.
Moreover the  $M_{gas}$ values obtained by Model 4 in the last 9 Gyr are higher due to the delayed
contribution of LIMS. For comparisons between closed box models with and without IRA see
Fig. 12 in Prantzos (2007).

Despite the fact that it is known that the [Fe/H] distribution shown in dwarf stars of the solar 
vicinity is impossible to reproduce with a closed box model, we use models 1--4
to check our approximation in C/O-O/H and $Y(O)$ relations.

  \subsection{Evolution of C/O with O/H}
  In Fig. 4, we show the evolution C/O-O/H of the models obtained with IRA, CHEVO and our delayed 
approximation.

Comparing models with the observational data, we conclude:
i) Model 3, assuming the canonical IRA for MS and LIMS, produces C/O values that are almost 
constant and  much higher than the observed ones for most of the evolution.
ii) Model 2, assuming IRA only by MS, reproduces very well the C/O evolution until 12 + 
log(O/H) $\sim 8$ because the ISM in the halo formation is determined mainly by MS.
Also, this model shows the C/O rise due to the high C yield of MS at high $Z$. 
However, this increase 
is not to enough to reach the high C/O values present in disk stars and in HII regions.
iii) Model 1, assuming the delayed contribution of LIMS, reproduces quite well the whole 
C/O-O/H evolution and specifically the increase of C/O abundance ratio at high O/H.

It is notable that these results are very similar to  results obtained with the 
numerical code that considers the lifetime 
of each star (model 4).

 \begin{figure}[!t]
  \includegraphics[width=\columnwidth]{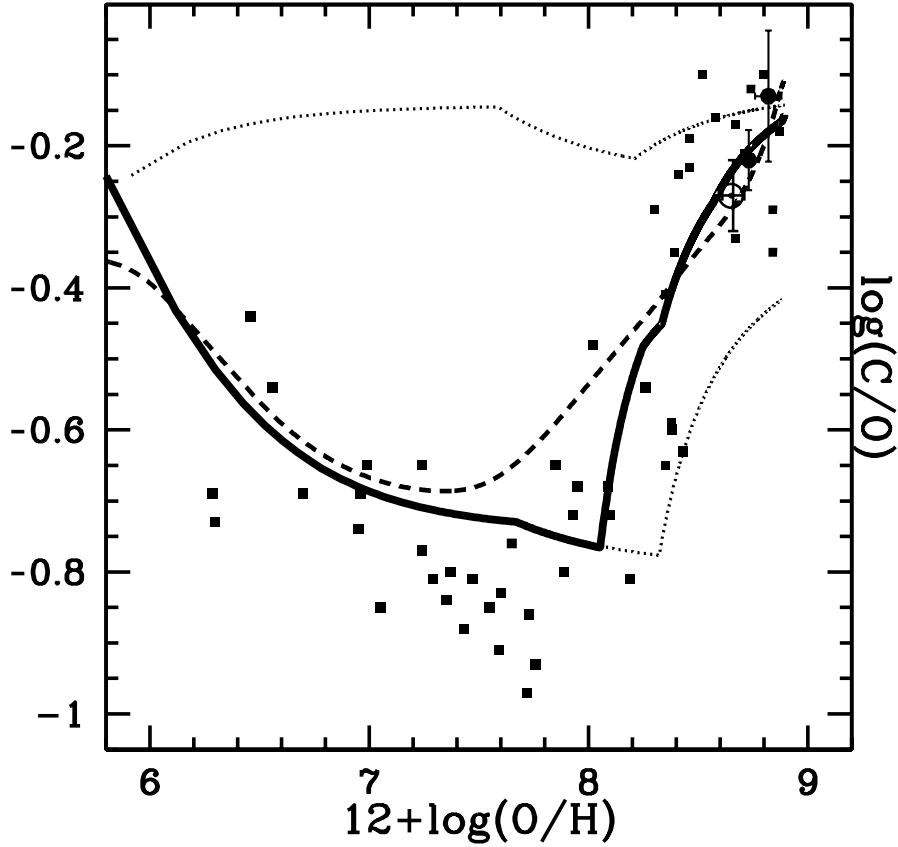}
  \caption{Evolution of C/O versus O/H using the different models:
{\it continuous line:} the delayed approximation,
{\it dashed line:} taking into account the lifetime of each star,
{\it lower dotted line:} IRA with MS,
{\it upper dotted line:} IRA with MS+LIMS,
   Observations: {\it circles:} HII regions by Esteban et al. (2005)
and Garc\'{\i}a-Rojas et al. (2004),
{\it squares:} dwarf stars by Akerman et al. (2004),
{\it $\odot$}:  solar values by Asplund et al. (2005). }
  \label{cosv} 
  \end{figure}

%-----------------------------------------------------------------------------
\subsection{Evolution of Helium with Oxygen}

In Fig. 5, we show the evolution of He by mass ($Y$) and O by mass ($O$)
obtained again using the models with IRA, CHEVO, and our delayed approximation.
In order to present the power of our approximation independent of initial abundances
we have plotted $\Delta Y$ and $\Delta O$.

Since there is not good observational data for He in dwarf stars in the solar vicinity
and in the nearest HII regions,
we have plotted the $Y$ and $O$ values from HII regions in dwarf galaxies and M17, an
inner Galactic HII region,  as representative of the past 
and future of the solar vicinity, respectively.

The $Y-O$ relation obtained from models 1 and 4 keep an almost linear trend when $O < 2.5 \times 10^{-3}$
and $O < 4.0 \times 10^{-3}$, respectively,
with a similar $\Delta Y/\Delta O$. 
Then the linearity is lost in both models and the slopes are more pronounced
due  mainly to the $Z$ dependence of O yields.
The difference in the changing O
is caused by the interpolation assumed between consecutive $Z_{pop}$.
In CHEVO code a linear interpolation in stellar yields is used 
when $Z_{gas}$ increases with time, while in our delayed approximation
we assumed constant yields: the average yields between consecutive $Z_{pop}$.

Again the delayed approximation is better than IRA and is more similar to that obtained by 
the numerical code.

 \begin{figure}[!t]
 \includegraphics[width=\columnwidth]{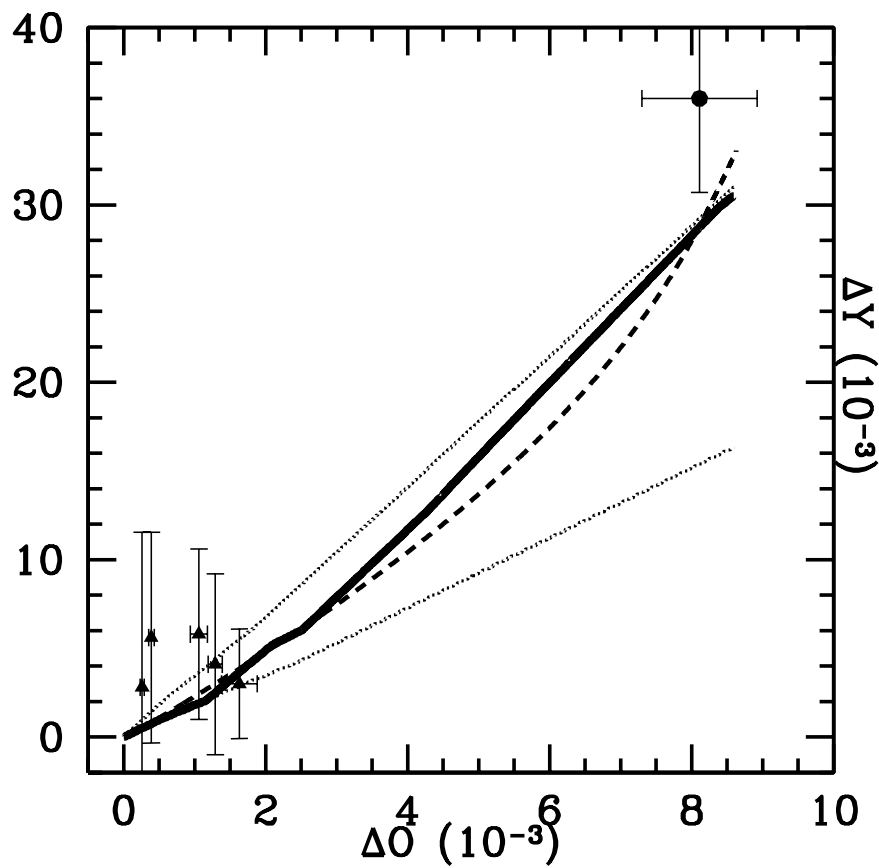}
 \caption{Evolution of Helium versus Oxygen by mass. Models as Fig. 4. 
{\it Observational data:}
M17, HII region at $r=6$ kpc by Carigi \& Peimbert (2008) ({\it filled circle}),
extragalactic HII regions by Peimbert et al. (2007) ({\it filled triangles}).}
  \label{yosv} 
  \end{figure}

\section{Conclusions}
\begin{itemize}

\item{We have found analytical equations for chemical evolution in the
case of a closed box model and SFR proportional to the gas mass where
the delayed enrichment by LIMS is represented by a single type of star.
}

\item{The delay of LIMS with respect to the galactic enrichment for He, C and N produces an
    increase on C/O and N/O with O/H and in Y with O in good agreement with the results obtained by
    numerical models. With IRA, the C/O and N/O values are constant with O/H in disagreement with 
    observed data and with  model results that take into account all star lifetimes.}

\item{For $\mu = 0.1$ and  $Z_{pop}=0.02$, $Y(O)$, C/O, and N/O values show artificial
     secondary raises  due to the O dilution produced by LIMS and not because of the 
    increase of He, C and N produced by the LIMS.}

\item{The delayed approximation was probed successfully in the solar vicinity reproducing the main 
trends of C/O-O/H relation shown by dwarf stars in agreement with results obtained with a numerical 
model that considers all star lifetimes. That relation cannot be reproduced at all by the 
instantaneous recycling approximation.}

\item{
The analytical equation (eq. 5) obtained by our approximation is a useful
 tool to know the chemical evolution of those
elements produced by LIMS when no galactic chemical evolutionary code
is available.
}

\end{itemize}

{\it Part of this work was submitted in the Physics Undergraduate Program at the
    Universidad Nacional Aut\'onoma de M\'exico.}

\begin{acknowledgements}
L. C. is grateful to a careful reading of the manuscript from A. E. Sansom.
I. F. thanks to E. Bell and A. Mart\'{\i}nez-Sansigre for helpful 
comments. 
We acknowledge the anonymous referee for excellent suggestions.
This work was partly supported by the CONACyT grants 46904 and 60354. 
\end{acknowledgements}


\begin{thebibliography}

\bibitem[Akerman et al.(2004)]{ake04}
Akerman, C. J., Carigi, L., Nissen, P. E., Pettini, M., \&
Asplund, M. 2004,
\aap, 414, 931

\bibitem[Asplund et~al.(2005)]{asp05}
Asplund, M., Grevesse, N., \& Sauval, A. J. 2005,
in: Cosmic Abundances as Records of Stellar Evolution and
Nucleosynthesis, ed. F. N. Bash \& T. G. Barnes, ASP Conference
Series, 336, 25

\bibitem[Carigi (1994)]{car04}
Carigi L.,  1994, ApJ, 424, 181

\bibitem[Carigi \& Peimbert (2008)]{car08}
        Carigi, L. \& Peimbert, M. 2008,
Rev. Mex. Astron. Astrof.,  submitted (arXiv:0801.2867)

\bibitem[\protect\citeauthoryear{Esteban et al.}{1998}]{est98}
Esteban, C., Peimbert, M., Torres-Peimbert, S., \& Escalante, V. 1998,
\mnras, 295, 401


\bibitem[\protect\citeauthoryear{Esteban et al.}{2005}]{est05}
Esteban, C., Garc\'{\i}a-Rojas, J., Peimbert, M., Peimbert, A., Ruiz,
M. T.,
Rodr\'{\i}guez, M., \& Carigi, L. 2005,
\apj, 618, L95


\bibitem[Garc\'{\i}a-Rojas et al.(2004)]{gar04}
Garc\'{\i}a-Rojas, J., Esteban, C., Peimbert, M., Rodr\'{\i}guez, M., Ruiz, M. T., \& Peimbert, A. 2004,
\apjs, 153, 501

\bibitem[Hirschi(2007)]{hir07}
Hirschi, R. 2007,
\aap, 461, 571

\bibitem[Kroupa, Tout \& Gilmore (1993)]{ktg93}
Kroupa, P., Tout, C. A., \& Gilmore, G. 1993,
\mnras, 262, 545

\bibitem[Maeder(1992)]{mae92}
Maeder,  A. 1992,
\aap, 264, 105

\bibitem[Meynet \& Maeder(2002)]{mey02}
Meynet G. \& Maeder A. 2002,
\aap, 390, 561

\bibitem[Pagel(1989)]{pag89}
Pagel B.E.J. 1989,
Rev. Mex. Astron. Astrof., 18, 161

\bibitem[\protect\citeauthoryear{Peimbert, Luridiana, \& Peimbert}
{Peimbert et~al.}{2007}]{pei07}
Peimbert, M., Luridiana, V., \& Peimbert, A. 2007,
\apj, 666, 636

\bibitem[Prantzos(2007)]{pra07}
Prantzos, N. 2007,
 in: ``Stellar Nucleosynthesis: 50 years after B2FH", 
eds. C. Charbonnel \& J.P. Zahn, EAS publications Series
(arXiv:0709.0833)


\bibitem[Talbot \& Arnett (1971)]{tal1971}
Talbot, R. Jr. \& Arnett, W. D. 1971,
\apj, 170, 409 

\bibitem[Tinsley (1974)]{tin74}
Tinsley, B.M.  1974,
ApJ, 192, 629

\bibitem[Schaller (1992)]{sch92}
Schaller, G., et al. 1992,
A\&AS, 96, 269

\bibitem[Serrano \& Peimbert (1983)]{ser83}
Serrano, A. \& Peimbert, M. 1983,
Rev. Mex. Astron. Astrof., 8, 117


\bibitem[van der Hoek (1997)]{van97}
van der Hoek, L. B. \& Groenewegen, M. A. T. 1997,
\aaps, 123, 305

\end{thebibliography}
\end{document}